\documentclass[prd,letterpaper,twocolumn,preprintnumbers,nofootinbib]{revtex4}

\usepackage{amsmath,amssymb}
\usepackage{graphicx}
\usepackage{units}
\usepackage{slashed}
\usepackage[hyperfootnotes=false,colorlinks,citecolor=blue]{hyperref}
\usepackage{setspace}
\newcommand{\beq}{\begin{equation}}
\newcommand{\eeq}{\end{equation}}
\newcommand{\bea}{\begin{eqnarray}}
\newcommand{\eea}{\end{eqnarray}}

\def \epsilon {\varepsilon} %different epsilon symbol

\begin{document}
%%%%%%%%%%%%%%%%%%%%%%%%%%%%%%%%%

\title{Type III seesaw with R-parity violation in light of $m_W$ (CDF)}

\author{Anish Ghoshal\textsuperscript{1}, Nobuchika Okada\textsuperscript{2}, Satomi Okada\textsuperscript{3}, Digesh Raut\textsuperscript{4}, Qaisar Shafi\textsuperscript{5}, Anil Thapa\textsuperscript{6}    \endgraf
\itshape \textsuperscript{1}Institute of Theoretical Physics, Faculty of Physics, University of Warsaw, ul. Pasteura 5, 02-093 Warsaw, Poland\\
\textsuperscript{2,3}Department of Physics and Astronomy, University of Alabama,
Tuscaloosa, Alabama, 35487, USA \\
\quad \textsuperscript{4}Department of Physics, Washington College,
Chestertown, Maryland, 21620, USA\\
\textsuperscript{5}Department of Physics and Astronomy, University of Delaware,
Newark, Delaware, 19716, USA\\
\textsuperscript{6}Department of Physics, University of Virginia,
Charlottesville, Virginia 22904-4714, USA
}

\email{anish.ghoshal@fuw.edu.pl, okadan@ua.edu, satomi.okada@ua.edu, draut2@washcoll.edu, qshafi@udel.edu, wtd8kz@virginia.edu}

%%%%%%%%%%%%%%%%%%%%%%%%%%%%%%%%%%
%%%%%%%%%%%%%%%%%%%%%%%%%%%%%%%%%%

\begin{abstract}
Motivated by the recently reported measurement of the $W$ boson mass $M_W = 80.4335 \pm 0.0094$ GeV by the CDF collaboration, we propose a type III seesaw extension of the minimal supersymmetric standard model (MSSM) which also includes an R-parity violating term. Without taking potential SUSY radiative corrections into account, we show that the CDF measurement of $M_W$ and the LEP measurement of the $\rho$ parameter can be simultaneously accommodated at the $2 \sigma$ level. A long-lived gravitino in a few GeV mass range is a unique viable dark matter candidate in this framework.
\end{abstract}

%%%%%%%%%%%%%%%%%%%%%%%%%%%%%%%%%
%%%%%%%%%%%%%%%%%%%%%%%%%%%%%%%%%
\maketitle

\textbf{\textit{Introduction:}} The CDF collaboration at Fermilab has reported a very precise measurement of the $W$ boson pole mass $m_W^{\rm CDF} = 80.4335 \pm 0.0094$ GeV \cite{CDF:2022hxs}, which shows about $7 \sigma$ deviation from the Standard Model (SM) prediction, $m_W^{\rm SM} = 80.357 \pm 0.006$ GeV \cite{Awramik:2003rn}. 
For the central values, the deviation is $\Delta m_W = m_W^{\rm CDF} - m_W^{\rm SM} = 0.0765$ GeV, which may be viewed as tantalizing evidence for the new physics beyond the SM. Following the CDF report, several works have examined new physics scenarios to account for the observed deviation \cite{Lu:2022bgw,Strumia:2022qkt,deBlas:2022hdk,Fan:2022yly,Tang:2022pxh,Cacciapaglia:2022xih,Blennow:2022yfm,Liu:2022jdq,Lee:2022nqz,Cheng:2022jyi,Song:2022xts,Bagnaschi:2022whn,Paul:2022dds,Bahl:2022xzi,Asadi:2022xiy,DiLuzio:2022xns,Athron:2022isz,Gu:2022htv,Babu:2022pdn,Heo:2022dey,Crivellin:2022fdf,Endo:2022kiw,Biekotter:2022abc,Balkin:2022glu,Han:2022juu, Du:2022pbp,Du:2022brr,Sakurai:2022hwh,Arias-Aragon:2022ats,Athron:2022qpo,Heckman:2022the}.

In addition to the CDF results, one should also consider the electroweak precision measurements (EWPM) by the LEP experiment. 
With the very precisely measured $Z$ boson mass $m_Z$ by the LEP experiment \cite{Electroweak:2003ram}, 
the SM prediction for the $\rho$-parameter is consistent with the LEP result, $\rho \equiv \frac{m_W^2}{ m_Z^2 c_W^2} = 1.0003 \pm 0.0005$ \cite{Workman:2022ynf}. 
Hence, a large deviation of $W$ boson mass from its SM predicted value is likely in tension with the LEP result. 
The challenge for any new physics scenario is to minimize the tension between the two results. 

In this letter, we provide a simple extension of the MSSM which incorporates type III seesaw and which can accommodate to a large extent the $\Delta m_W$ in the presence of a suitable R-parity violating term. The model is also consistent with the LEP results and satisfies various other phenomenological constraints, in addition to providing a framework for neutrino masses and mixings. With the inclusion of the gauge-mediated supersymmetry breaking, the model can also accommodate successful inflationary scenario. Finally, the model can be embedded within grand unified theory (GUT) such as $SU(5)$.\\

\begin{figure}[!t]
\centering{
   \includegraphics[scale=0.25]{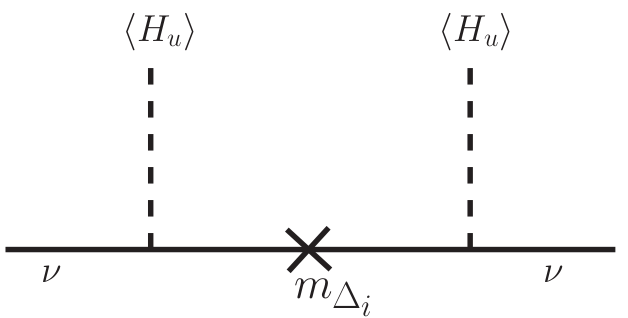} ~~~~~~~~~~~~~~~~~~~~~~~~~~~~~~~~~~~~(a) \\
   \vspace{3mm}
    \includegraphics[scale=0.22]{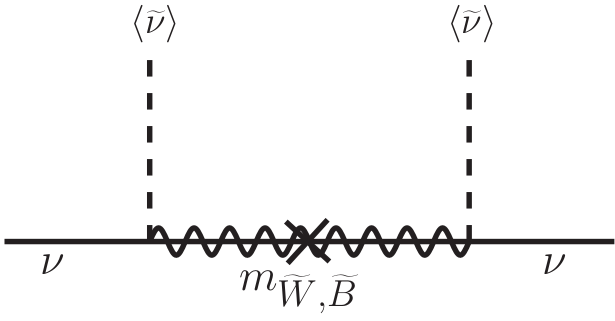}\\
    (b)
    \caption{\it Neutrino mass generation mechanism in the model.}
    \label{fig:numass}
    }
\end{figure}

\textbf{\textit{The Model:}} MSSM is a well-motivated candidate for physics beyond the SM that provides a natural resolution for the SM gauge hierarchy problem ~\cite{Drees:2004jm,Baer:2006rs,Aitchison:2005cf,Rodriguez:2016esw,Rodriguez:2019mwf}. However, it cannot explain the origin of the observed neutrino mass as shown by the solar and atmospheric neutrino oscillation experiments \cite{Electroweak:2003ram}. 
To remedy this, we implement type III seesaw by including two $SU(2)_L$ triplet chiral `matter' superfields, $\Delta_i$ ($i = 1, 2$) with zero hypercharge. 
The MSSM superpotential is extended to include 
\bea
W \supset \sum_{i=1}^{2} \sum_{j=1}^{3} \sqrt{2} Y_{D}^{i j} H_{u}^{T} \varepsilon \Delta_{i} L_{j}-\frac{m_{\Delta}}{2}  \sum_{i=1}^{2} Tr\left[\Delta_{i} \Delta_{i}\right]. 
\label{eq:pot}
\eea
Here, $H_u$ and $L_i$ respectively denote the up-type Higgs doublet and the lepton doublet, and 
\bea
\varepsilon=\left(\begin{array}{cc}
0 & 1 \\
-1 & 0
\end{array}\right), 
\qquad 
\Delta_{i}=\frac{1}{\sqrt{2}}\left(\begin{array}{cc}
\Delta_{i}^{0} & \sqrt{2} \Delta_{i}^{+} \\
\sqrt{2} \Delta_{i}^{-} & -\Delta_{i}^{0}
\end{array}\right). 
\eea
The Higgs doublets, $H_{u,d}$ develop nonzero vacuum expectation value (VEVs) in the usual way, which break the electroweak symmetry, $\langle H_u\rangle = \left(0, v_u/ \sqrt{2}\right)^T $ and $\langle H_d\rangle = \left( v_d/\sqrt{2},0\right)^T $, where $v_u = v \cos\beta$, $v_d = v \sin\beta$ and $ v = \sqrt{v_u^2 + v_d^2} = 246$ GeV, and $H_d$ is the down-type Higgs doublet.  
Assuming $m_\Delta \gg Y_D v_u$, the light Majorana neutrino mass matrix generated via the type III seesaw mechanism shown in Fig.~\ref{fig:numass} (a) is given by 
\bea
m_{\nu}=\frac{Y_{D}^{T} Y_{D} v_{u}^{2}}{2 m_{\Delta}} \sim \frac{Y_{D}^{T} Y_{D} v^{2}}{2 m_{\Delta}}, 
\eea
for $\tan\beta = v_u/v_d  \gtrsim 10$. 

We also introduce the R-parity (lepton-number) violating term in the superpotential, 
\bea
W_{\not R} =\sqrt{2} \lambda H_{u}^{T} \varepsilon \Delta_{1} H_{d}. 
\label{eq:RVT1}
\eea
Note that we have only used $\Delta_1$ to break R-parity for simplicity. The inclusion of this term is crucial to generate an induced VEV for ${\widetilde \Delta_1} $ (the scalar component of $\Delta_1$). 
This helps increase the $W$ boson mass at tree-level, without altering the $Z$ boson mass, which is consistent with the LEP result. To illustrate this, we consider the  trilinear scalar soft SUSY breaking term corresponding to $W_{\not R}$ which is of the following form:   
\bea
V \supset \sqrt{2} \lambda A H_{u}^{T} \varepsilon \widetilde{\Delta} H_{d}+h. c., 
\eea
where the parameter $A$ has mass dimension of one. After the electroweak symmetry breaking, and taking into account the mass squared term for $\widetilde{\Delta}_1^0$ in Eq.~\ref{eq:pot}, the relevant potential is given by 
\bea
V \supset m_{\Delta}^2 |\widetilde{\Delta}_1^0|^2 -\frac{1}{2} \lambda A v_{u} v_d  \widetilde{\Delta}_1^0 + h.c. .
\eea
As a result $\widetilde{\Delta}_1^0$ acquires an induced VEV, 
\bea
\left\langle\widetilde{\Delta}_{1}^{0}\right\rangle=\frac{1}{2} \lambda A v_{u} v_{d} / m_{\Delta}^{2} \equiv v_{\Delta} / \sqrt{2} \, .
\eea
Here, we have neglected the soft SUSY breaking mass-squared term for $\widetilde{\Delta}_1^0$ by assuming it to be much smaller than $m_\Delta^2$. 
With $v_\Delta \neq 0$, the $W$ and $Z$ masses are given by
\bea
m_{W}^{2}&=&\frac{g^{2}}{4}\left(v^{2}+4 v_{\Delta}^{2}\right) \, , \\
m_{Z}^{2}&=&\frac{g^{2}}{4 \cos \theta_W^{2}} v^{2} \, , 
\eea
where $\cos\theta_W$ is the electroweak mixing angle. 
Clearly, since $\Delta_1$ has zero hypercharge, it only contributes to the $W$ boson mass. 
For $v_\Delta^2 \ll v^2$, 
\bea
\Delta m_W = m_W - m_W^{\rm SM} \simeq 2 m_W^{\rm SM} \left( \frac{v_\Delta^2}{v^2}\right) \, .
\label{eq:delW}
\eea
To reproduce the CDF measurement within $n-\sigma$ of the central value, we require 
\bea
v_\Delta^{\rm CDF} ({\rm GeV}) = 19.4 \times \sqrt{0.0765-n \times 0.0094}. 
\label{eq:vdelta}
\eea
%%%%%
As previously discussed, we should also consider the LEP results for EWPM, in particular the $\rho$ parameter. 
At tree-level it is given by  
\bea
\rho \equiv \frac{ m_W^2}{ m_Z^2 c_W^2} = 1 + 4\left( \frac{v_\Delta^2}{v^2}\right) \equiv 1 + \Delta \rho \, .
\label{eq:rho}
\eea
To reproduce the LEP measurement $\rho = 1.0004 \pm 0.0005$ \cite{Electroweak:2003ram, ParticleDataGroup:2020ssz} within $n-\sigma$ of the central value requires,\footnote{
Essentially the same strategy is proposed in Ref.~\cite{FileviezPerez:2022lxp} to reproduce the CDF measurement, where an $SU(2)_L$ triplet scalar with zero hypercharge
and its VEV are introduced. 
Our main proposal in the present paper is that such a triplet scalar is automatically implemented in the type III seesaw extension of the MSSM
for neutrino mass generation and the scalar VEV is induced by the R-parity violating term. 
}
\bea
v_\Delta^{\rm LEP} ({\rm GeV}) = 123.4 \times \sqrt{0.0004+n \times 0.0005} \, .
\label{eq:vrho}
\eea
Clearly, the CDF and LEP results are in tension with each other, which is indicated by the fact that  Eqs.~(\ref{eq:vrho}) and (\ref{eq:vdelta}) are inconsistent with each other for the central values.  
In our analysis, we only consider the tree-level effect from the triplet scalar. 
See Refs.~\cite{Forshaw:2001xq,Forshaw:2003kh} for a detailed analysis of oblique corrections and global fit to EWPM
in the presence of a triplet scalar field. 
It has been shown that for a triplet mass $\gtrsim$ 1 TeV, the oblique corrections have negligible effects for the fit. 

\begin{figure}
    \centering{
    \includegraphics[width=0.49\textwidth]{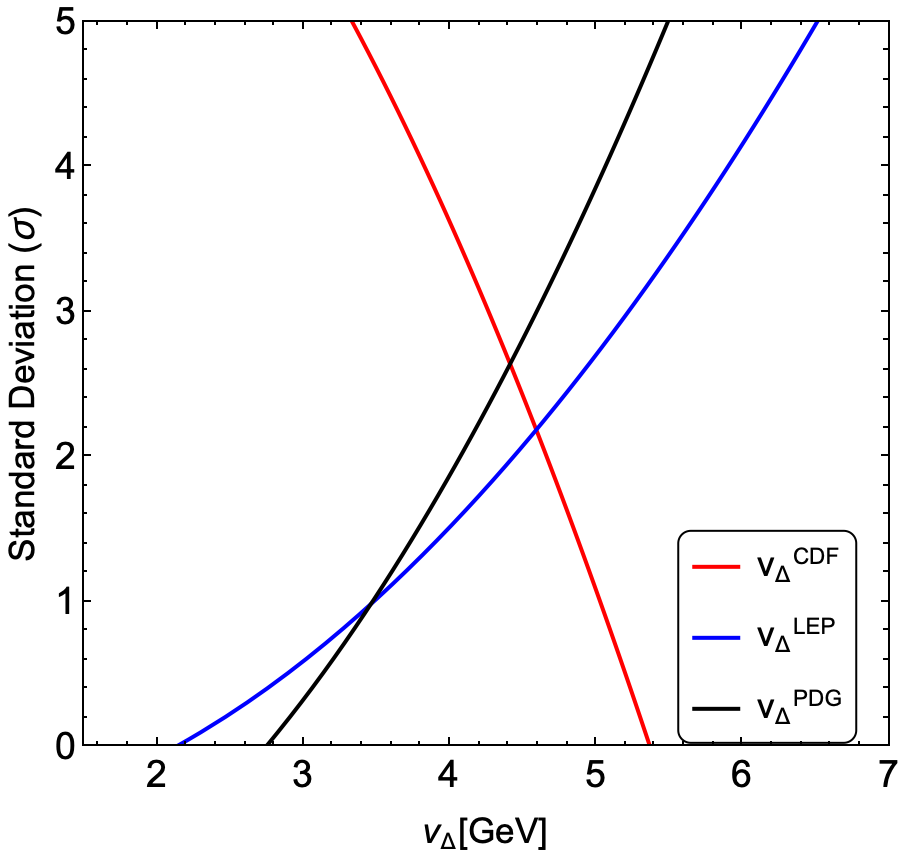}
    \caption{\it Induced {\rm VEV} $v_\Delta^{\rm CDF}$ (red), $v_\Delta^{\rm LEP}$ (blue), and $v_\Delta^{\rm PDG}$ (black) versus n$-\sigma$ deviation.}
    \label{fig:plot}
    }
\end{figure}
In Fig. \ref{fig:plot}, we show a plot of $v^{\rm CDF}_{\Delta}$ (GeV) and 
$v^{\rm LEP}_{\Delta}$ (GeV) versus the number of standard deviations $n$. 
One needs at least 2$\sigma$ deviations for both $m^{\rm CDF}_{W}$ and $\Delta \rho$ to be consistent with each other. 
For comparison, in addition to LEP and CDF-II results, we also show a line (black) corresponding to the world average of 
$M_W^\textrm{PDG}= 80.377 \pm 0.012 \;\textrm{GeV}$~\cite{ParticleDataGroup:2022pth}, 
which takes into account the $W$ mass measurements from LEP~\cite{ALEPH:2013dgf}, Tevatron~\cite{CDF:2013dpa} (CDF~\cite{CDF:2012gpf} and D0~\cite{D0:2012kms}) and LHCb collaboration~\cite{LHCb:2021abm}.
The dependence of standard deviations $n$ on $v_\Delta^{\rm PDG}$ is similar to that of $v_\Delta^{\rm LEP}$ and
prefers smaller values of the induced VEV. 

We therefore conclude that in the type III seesaw extension of MSSM, if an R-parity violating term is introduced, the precisely determined $m^{\rm CDF}_{W}$ can be accommodated in the model with $v_{\Delta} \sim 4$ GeV, which is induced by the appearance of the soft-SUSY breaking trilinear scalar coupling term involving the type III seesaw messenger $\Delta_{i}$.\\

%%%%%%%%%%%%%%%%%%%
%%%%%%%%%%%%%%%%%%%
%%%%%%%%%%%%%%%%%%%%

\textbf{\textit{Constraints on R-parity Violation:}} To check the consistency of the model, let us next consider the phenomenological constraints on R-parity violation. 
We first note that $v_\Delta$ generates the so-called bi-linear R-parity violating term: 
\bea
W \supset \sqrt{2}\left(Y_{b}\right)^{1 j} H_{u}^{T} \varepsilon\left\langle\widetilde{\Delta}_{1}\right\rangle L_{j} \equiv \mu_{\Delta}^{j} H_{u}^{T} \sigma^{1} L_{j} \, ,
\eea
where $\sigma$ is the Pauli matrix. 
Together with the MSSM $\mu$-term, 
\bea
W \supset \mu_{H} H_{u}^{T} \varepsilon H_{d} \, , 
\eea
the $H_d$ and $L_j$ fields mix, and the mixing angles are 
\bea
\varepsilon_{\hat{j}} \sim \frac{\mu_{\Delta}^{j}}{ \mu_{H}} \sim\left(Y_{D}\right)^{1 {j}} \frac{v_{\Delta}}  {\mu_{H}} \, .
\eea 
This leads to R-parity violating terms (lepton number violating Yukawa interactions):  
\bea
W_{\not R} \supset Y_{e}^{i j} \!E_{i}^{c} L_{j}\!\Big(\!\sum_{k} \varepsilon_{k} L_{k}\!\Big) + Y_{d}^{i j} \!D_{i}^{c} Q_{j}\!\Big(\!\sum_{k} \varepsilon_{k} L_{k}\!\Big) 
\eea
Such lepton number violating processes can be active in the early universe and wash out the baryon asymmetry. 
To avoid such a wash-out, we must ensure that these lepton number changing processes fall out of equilibrium for temperature $T\gtrsim {\tilde m}$, where ${\tilde m}$ is a typical sparticle mass. For ${\tilde m} = {\cal O} (1 {\rm TeV})$, 
this requires (see Refs.~\cite{Christodoulakis:1990tua, Fischler:1990gn, Dreiner:1992vm, Burell:2011wh})
\bea
\left(Y_{e, d}\right)^{i j} \varepsilon_{k} \lesssim 10^{-7}. 
\eea
For the Yukawa couplings to be in perturbative regime, the above conditions is always satisfied if 
\bea
\varepsilon_{k} \sim \frac{\left(Y_{D}\right)^{1 k} v_{\Delta}}{\mu_{H}} \lesssim 10^{-7}. 
\eea
We have shown that $v_{\Delta} = {\cal O} (1 {\rm GeV})$ is required to explain the CDF measurement of $W$ boson mass, which naturally leads to $\left(Y_D\right)^{1k} \lesssim 10^{-4}$ for $\mu_H \sim {\tilde m} = {\cal O} (1 {\rm TeV})$. Thus, using the type III seesaw formula, one can roughly estimate the size of $\left(Y_D\right)^{1k}$ as 
\bea
\left(Y_{D}\right)^{1 k} \sim \frac{\sqrt{m_{\nu} m_{\Delta}}}{v_{u}}, 
\eea
where $m_\nu \sim 10^{-10}$ GeV is the neutrino mass scale and $v_u \sim v = 246$ GeV is the electroweak VEV. 
We find that the bound on the Yukawa coupling $\left(Y_D\right)^{1k} \lesssim 10^{-4}$ is satisfied for
\bea
m_\Delta \lesssim 6\times 10^{6}\ {\rm GeV} .
\eea
%%%
It is worth pointing out that the left-handed sneutrinos ($\widetilde{\nu}_i$) in the lepton doublet superfields can develop induced VEVs, $\widetilde{\nu}_i \sim \varepsilon_i v_d$, through their mixing with $H_d$. 
This generates neutrino masses, shown in Fig.~\ref{fig:numass} (b), and the mass given by 
\bea
m_{\nu} \sim \frac{g^{2} \langle \widetilde{\nu}_i \rangle^{2}}  { m_{\widetilde{W}}}, 
\eea
where $m_{\widetilde{W}}$ is the wino mass. 
For $m_{\widetilde W} = {\widetilde m} \simeq {\cal O}(1 {\rm TeV})$ and $\varepsilon_i \lesssim 10^{-7}$, we obtain 
\bea
m_{\nu}^{\text {new }} \sim \frac{\left(10^{-7} v_{d}\right)^{2}}{ m_{\widetilde{W}}}  \sim 10^{-13}\ \mathrm{GeV}, 
\eea
which is much smaller than the typical neutrino mass scale of $10^{-10}$ GeV required by the neutrino oscillation data \cite{Esteban:2020cvm}. 
Therefore, the R-parity violation has no significant effect on the type III seesaw mechanism.\\

\textbf{\textit{Gravitino Dark Matter:}}
\begin{figure}
    \centering
    \includegraphics[scale=0.23]{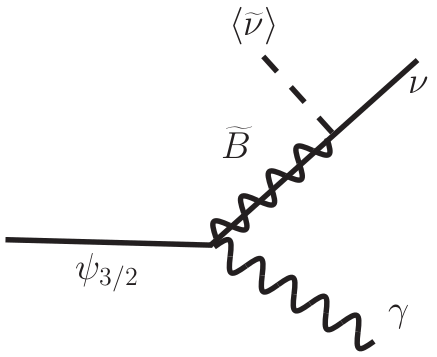}
    \caption{\it Gravitino DM decay into neutrino and a photon ($\nu + \gamma$).}
    \label{fig:graDM}
\end{figure}
In MSSM, a stable neutralino is a standard WIMP candidate for DM. 
Since R-parity is violated in our model, the neutralino is not stable and hence it is not a viable DM candidate.  
We show how a long-lived gravitino 
is a viable DM candidate. 
Let us consider a gravitino with mass in the few GeV range which mainly decays to $\gamma + \nu$ \cite{ Takayama:2000uz} as shown in the Fig.~\ref{fig:graDM}.
Its decay width is approximately given by 
\bea
\Gamma_{\psi_{3 / 2} \rightarrow \gamma \nu} \sim \frac{1}{16 \pi m_{3 / 2}} \left|\frac{(m_{3/2})^3}{\langle F\rangle}\right|^{2}\left|\frac{g_{Y}\langle\tilde{\nu}_i\rangle}{m_{\tilde{B}}}\right|^{2}, 
\eea
where $\langle F\rangle$ is the SUSY breaking order parameter, $m_{3/2} \sim \langle F\rangle/M_p$ is the gravitino mass, and $M_p = 2.44\times10^{18}$ is the reduced Planck mass. 
With $m_\nu^{\rm new} = g_{Y}\langle\tilde{\nu}\rangle \sim 10^{-13}$ GeV, $m_{3/2} = {\cal O} (1)$ GeV and $m_{\tilde{B}} = {\cal O} (1)$ TeV, we obtain 
$\tau_{3/2} \sim 10^{30}$ seconds for the gravitino lifetime, which makes it a viable candidate for (non-thermal)
cold DM.
For a discussion of gravitino DM scenarios in the context of type III seesaw extended MSSM, see Ref.~\cite{Mohapatra:2008wx}. 
In the paper, the type III seesaw extended MSSM is embedded into the SU(5) GUT framework and
the gauge mediated SUSY breaking scenario is unified with the type III seesaw mechanism
by identifying the type III seesaw messengers (the $SU(2)_L$ triplet chiral superfields) with the messenger fields
in the gauge mediated SUSY breaking scenario. 
Furthermore, it has been shown in Ref.~\cite{Kawai:2021gap} that a successful inflation scenario can be implemented to this model. 
Since the inclusion of the R-parity violating term does not alter the underlying physics of these scenarios,
we can realize both of these scenarios in our model.

\textbf{\textit{Conclusion:}} 
To summarize, we have presented an extension of MSSM which incorporates type III neutrino seesaw and also includes an appropriate R-parity violating term. This leads to a tree-level correction that increases the W boson mass, such that a 2-$\sigma$ agreement both with the recent determination of $m_W$ by CDF as well as the LEP measurement of the SM $\rho$ parameter is achieved (see Fig.~\ref{fig:plot}). 
Based on some recent papers, we expect to be able to reach the central CDF value for $m_W$ by taking into account radiative corrections involving some of the supersymmetric particles.
Since R-parity is not conserved, we identified a few GeV long-lived gravitino as a plausible candidate for (non-thermal) cold dark matter. Finally, our model can be readily embedded in a grand unified framework such as $SU(5)$.\\

\textbf{\textit{Acknowledgement:}} AT would like to thank Julian Heeck for the useful discussion. The work is supported by the United States Department of Energy grants DE-SC0012447 (N.O.) and DE-SC0013880 (Q.S.). 

\bibliographystyle{utcaps_mod}
\bibliography{BIB}

\end{document}